\documentclass[a4paper,11pt]{article}
\usepackage{jinstpub} % for details on the use of the package, please see the JINST-author-manual
\usepackage{lineno}
\title{\boldmath A Micromegas-based gaseous detector for  neutron-induced charged-particle reaction studies in nuclear astrophysics }

\author{C. Yadav,  A. Green, and M. Friedman\note{Corresponding author.}}
\affiliation{Racah Institute of Physics, The Hebrew University of Jerusalem, 91904, Israel}
\emailAdd{moshe.friedman@mail.huji.ac.il}

\abstract{ The quasistellar neutron spectrum produced via $^{7}$Li($p$, $n$)$^{7}$Be reaction at a proton energy of 1.912 MeV has been extensively studied and employed reaction for neutron-induced reaction studies.  We are working towards using this reaction at various proton energies from 1.9 MeV to 3.6 MeV to produce a neutron field at a temperature range of $\sim$1.5-3.5 GK to conduct measurements of neutron-induced charge particle reaction cross sections on various unstable nuclei at explosive stellar temperatures.  In this paper, we are reporting our design and simulation study with regards to experimental set-up and a gaseous detector with a segmented Micromegas detector for conducting neutron-induced charge particle reactions studies for nuclei of astrophysics importance, for example, $^{26}$Al($n$, $p$)$^{26}$Mg,  $^{26}$Al($n$, $\alpha$)$^{23}$Na and  $^{40}$K($n$, $p$)$^{40}$Ar,   $^{40}$K($n$, $\alpha$)$^{37}$Cl reactions. We plan to perform our experiments with a 10-$\mu$A proton beam at the Physikalisch Technische Bundesanstalt facility (PTB, Germany), with a Micromegas-based gaseous detector under construction as discussed in the paper.

 }


\arxivnumber{2308.12099} % Only if you have one

\begin{document}
\maketitle
\flushbottom

\section{Introduction}

Neutron-induced charged-particle reaction studies on various unstable nuclei are imperative for a complete understanding of the nucleosynthesis occurring in many astrophysical sites and events. Among them, a complete understanding of supernovae nucleosynthesis requires data on thousands of nuclear reaction rates of different types, most of them not yet experimentally accessible, at temperatures of 1.5-3.5 GK. Sensitivity studies have pointed out several ($n$, $p$) and ($n$, $\alpha$) reactions to having a relatively high impact on the final isotopic abundances  \cite{Illiadis, kh, hoff, wanjo, parikh}. Experimental cross-section data on ($n$, $p$) and ($n$, $\alpha$) reactions at the relevant energies for explosive nucleosynthesis is scarce or non-existent for most isotopes of interest. One reason is the need for an intense neutron source at energies of $\sim$10-2000 keV, corresponding to a Maxwellian flux distribution at temperatures of 1.5-3.5 GK. Another difficulty is the target size. Even for stable isotopes, the limited range of the outgoing charged particle imposes a strong limit on the effective target size. For unstable isotopes, the target size is an even bigger problem, both in terms of production and target handling. In Ref. \cite{MF} we discussed a novel approach to overcome those limitations by utilizing the $^{7}$Li($p$, $n$) reaction at varying proton energies in the range of 1.9-3.6 MeV on a thick lithium target to carry out neutron-induced charged-particle reactions cross-sections measurements in the temperature region of 1.5-3.5 GK.

As the activation approach is not feasible for most isotopes of interest, which naturally decay via electron capture or $\beta^+$ that are indistinguishable from the ($n$, $p$) reaction, prompt detection of the outgoing charged particle is essential. The prompt detection of charged particles is challenging in terms of the observed count rate and its detection approach in the overwhelming $\gamma$ and neutron fields. We have designed and performed a simulation study of a gaseous detector set-up for neutron-induced charged particle reaction cross-section measurement using $^{7}$Li($p$, $n$)$^{7}$Be reaction as a neutron source, to understand the challenges and feasibility of the experimental set-up. The details of the experimental approach,  detection method, and simulation study are discussed in detail in the following sections.

\label{sec:intro}

\section{The experimental set-up }

We plan to conduct neutron-induced charged-particle reaction studies at neutron energies of 1 - 2000 keV for nuclei important to supernovae nucleosynthesis with lifetimes on the order of 10$^5$ years or above, which we consider feasible with the current experimental setup. We plan to perform $^{26}$Al($n$, $p$)$^{26}$Mg,  $^{26}$Al($n$, $\alpha$)$^{23}$Na [t$_{1/2}$=7.2 x 10$^5$ Years], and $^{40}$K($n$, $p$)$^{40}$Ar,   $^{40}$K($n$, $\alpha$)$^{37}$Cl [t$_{1/2}$=1.25 x 10$^9$ Years] reactions cross-section measurements. Neutrons are produced via the $^{7}$Li($p$, $n$)$^{7}$Be reaction with a monoenergetic proton beam at various energies from 1.9 MeV - 3.6 MeV in a step of 0.1 MeV. The target of interest will be placed in a gaseous detector, and the outgoing charged particles are detected by ionization. 

\begin{figure}[ht!]
\centering
\includegraphics[width=16.2cm]{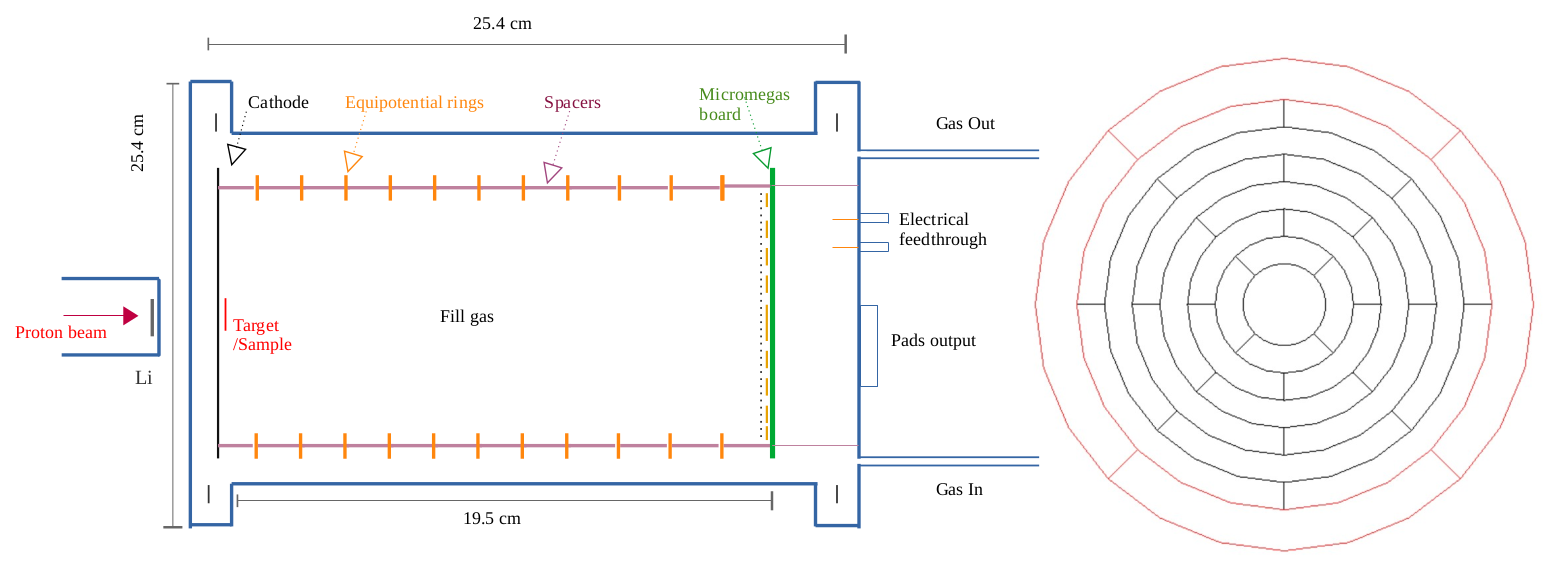}
\caption{(Left) Conceptual design of the experimental setup and a Micromegas-based gas-filled detector. (Right) Segmentation scheme of the anode pad.\label{fig:i}}
\end{figure}

\begin{figure}[ht!]
\centering
\includegraphics[width=9.0cm]{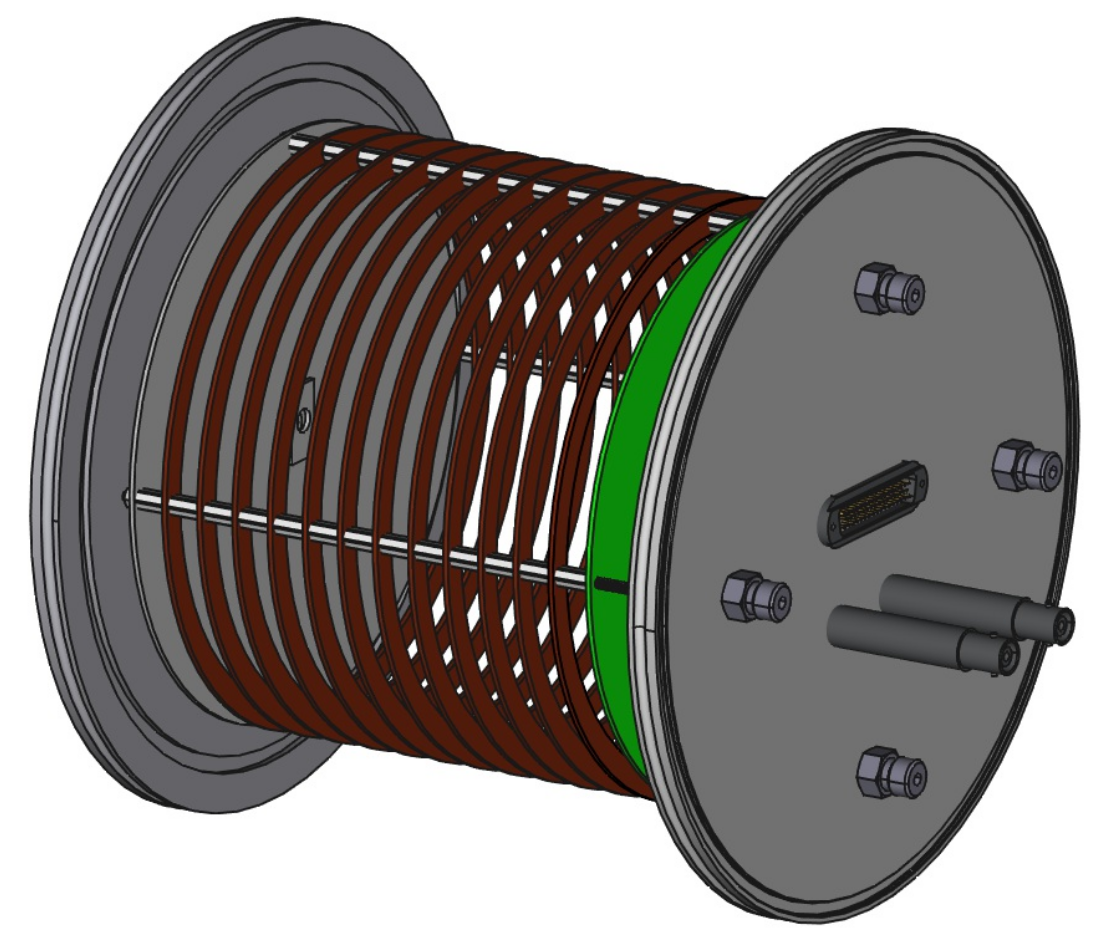}
\caption{Visualization of the detector set-up with CAD drawing showing components of the detector such as cathode, equipotential rings, spacers, anode plane, and feed-throughs as shown in the left panel of Fig.~\ref{fig:i}. \label{fig:i2}}
\end{figure}

\begin{figure}[ht!]
\centering
\includegraphics[width=10.0 cm]{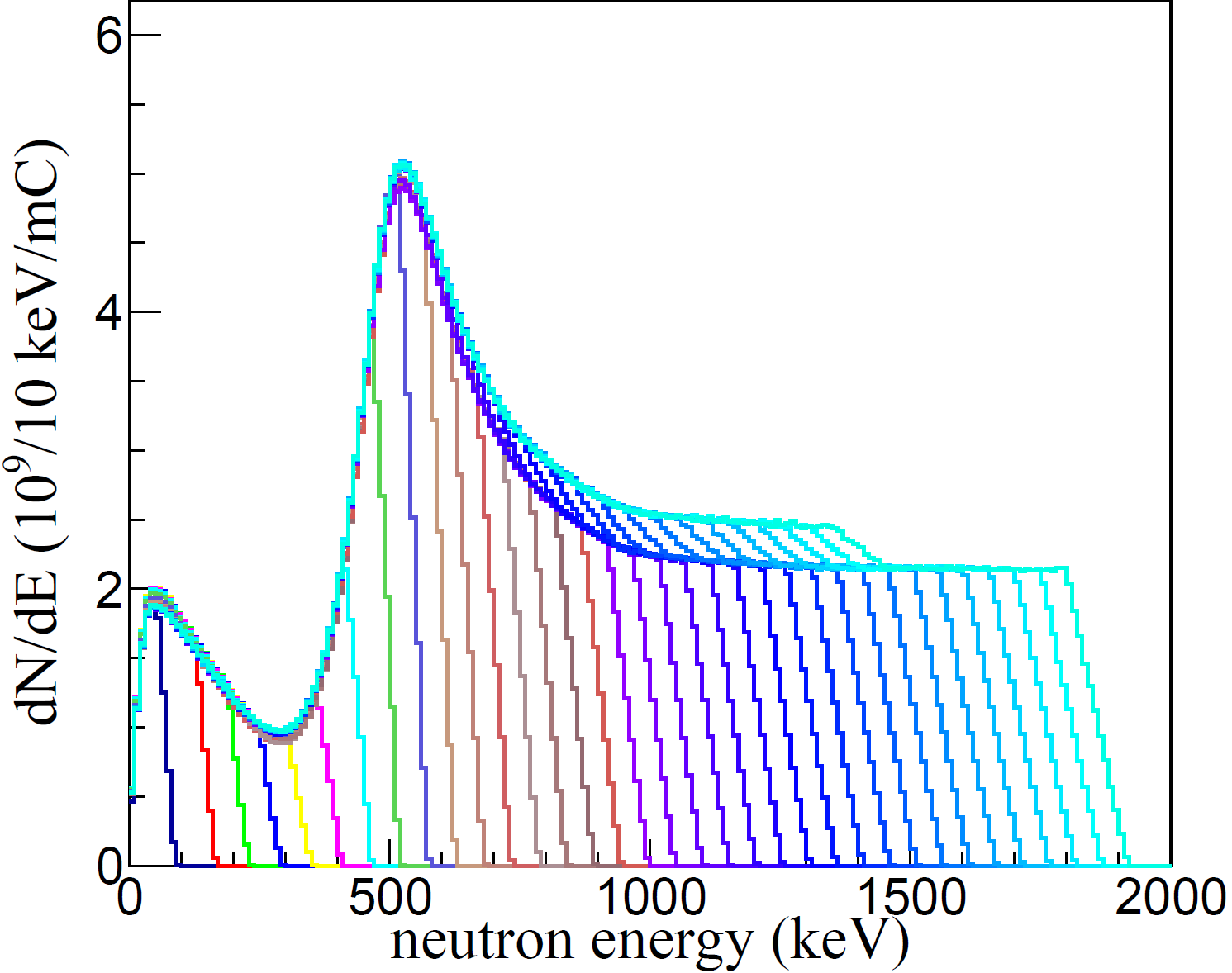}
\qquad
\includegraphics[width=10.0 cm]{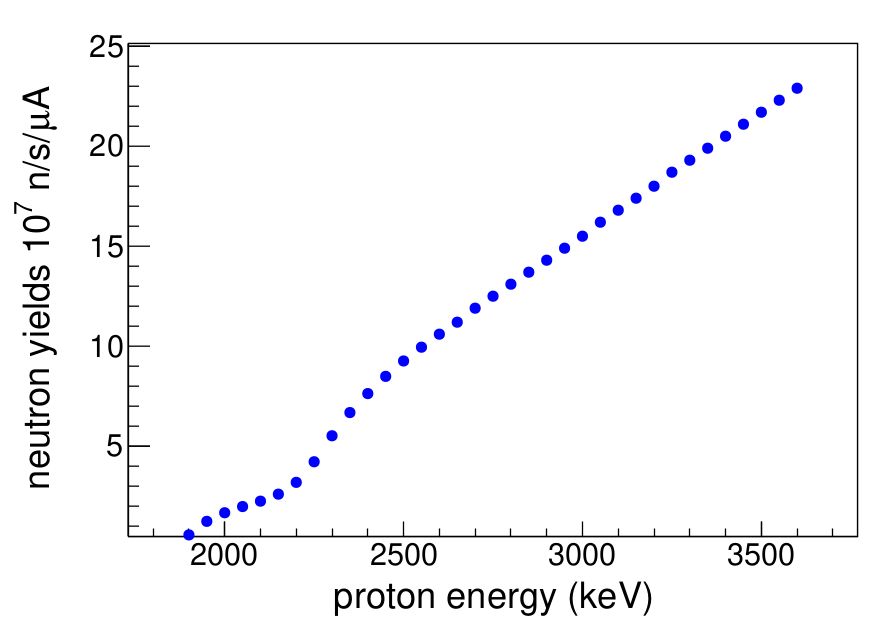}
\caption{(Top) Calculated neutron spectra at the target, for a thick lithium target, and a 5 cm$^2$ target located 2 cm downstream the lithium target. The different colors represent different proton beam energies from 1.9 MeV to 3.6 MeV. The neutron spectra were calculated using SimLiT  \cite{MF2}.     (Bottom) Total neutron intensity going through the target as a function of proton energy. The values are the integration over the spectra in the left figure. \label{fig:i3s}}
\end{figure}

\subsection{The experimental set-up and detector}
\label{sec-2}

The left panel of Fig.~\ref{fig:i} shows a schematic illustration of the experimental set-up and the right panel of Fig.~\ref{fig:i} shows the segmentation scheme of the Micromegas board.  Fig.~\ref{fig:i2} presents the CAD drawing of the detector showing the components of the detector such as cathode, equipotential rings, spacers, anode plane, and  feed-throughs as shown in the left panel of Fig.~\ref{fig:i}. As illustrated in the  left panel of Fig.~\ref{fig:i}, a monoenergetic proton beam with varying energies from 1.9 MeV - 3.6 MeV in a step of 0.1 MeV impinges on a thick Li target to produce a neutron beam via $^{7}$Li($p$, $n$)$^{7}$Be reaction, which further impinges on a target placed inside the Micromegas-based gaseous detector. The detector is located around 2 cm downstream of the neutron source.  Fig.~\ref{fig:i3s} (Top) shows the calculated neutron spectra at the target, for a thick lithium target, and a 5 cm$^2$ target located 2 cm downstream of the lithium target at proton energies from 1.9 MeV to 3.6 MeV. Fig.~\ref{fig:i3s} (Bottom) shows the total neutron intensity going through the sample as a function of proton energy. The neutron spectra were calculated using SimLiT  \cite{MF2}. Reaction products from the neutron-induced reaction are detected inside the gas-filled detector.

%\subsection{The detector }

The detector is cylindrical with a cathode, equipotential rings, and a segmented Micromegas detector. The sample will be placed in the center of the cathode inside the chamber facing the Micromegas. Such a design of the detection system provides a large solid angle for charged-particle detection. The separation between the cathode and the anode plane is around 19.5 cm, which acts as a drift gap between the cathode and Micromegas, and the active detection volume diameter is 18 cm. The segmentation scheme of the anode plane is shown in the right panel of Fig.~\ref{fig:i}.
The anode plane will be a segmented Micromegas detector.  It is divided into eight circular rings with a central pad ring of diameter 3 cm, second, third, fourth, fifth, sixth, and seventh rings of diameters,  5 cm, 7 cm, 9 cm, 11 cm, 13 cm, 15 cm, respectively, and eighth ring of diameter 18 cm. These outer rings are further segmented into four sections, thus giving 29 segmented pads on the anode plane, as shown in the right panel of Fig.1. The micromesh will be stretched over the anode and kept at a uniform distance of 128 $\mu$m  from the anode by placing insulating spacers every 5 mm. The principle of the detector is as follows. With the appropriate biases on the electrodes, the electrons created from the ionization of the filled gas (as described ahead) by particles emitted from the reaction in the active volume of the detector will drift towards the Micromegas and get amplified in the micromesh structure and thus get collected on the anode pads. The central pad with an active area of 7 $cm^2$ faces the target; hence it will detect any heavy charged particle emitted from the reaction. It will serve as a trigger to exclude most of the background events. The outermost ring, shown in red, will serve as a veto pad to exclude events that do not originate in the detector and particles that did not deposit all their energy in the active volume. The rest of the segmentation aims to allow particle identification based on energy loss 
(dE/dX) technique and reduce the background rates on each individual pad as discussed ahead.

\subsection{Fill gas }

\begin{figure}[ht!]
\centering
\includegraphics[width=13.3 cm]{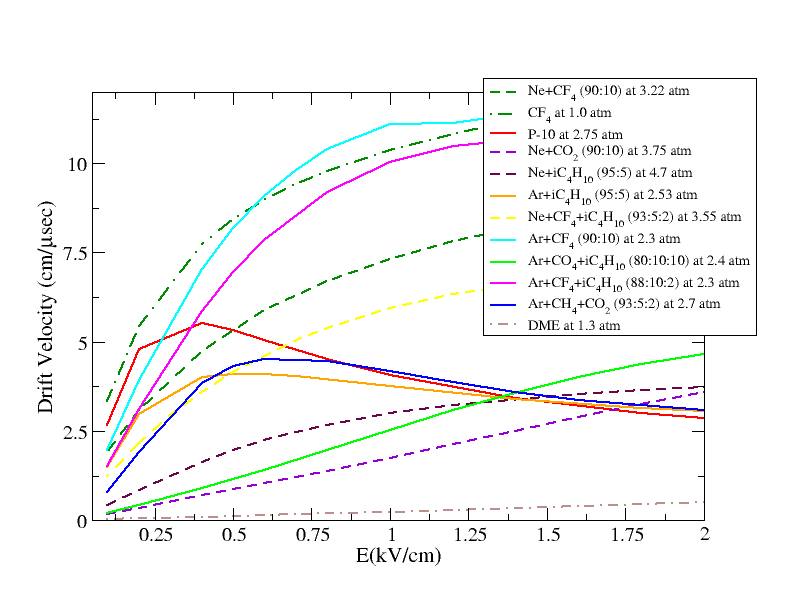}
\caption{Electron drift velocity as a function of the field for various gas mixtures obtained using \cite{mag}.\label{fig:i3}}
\end{figure}

We carried out a detailed analysis of various gas mixtures for the detector being developed for neutron-induced charge particle reaction experiments. The chamber gas pressure required to stop protons of energy up to 6.5 MeV from $^{26}$Al($n$, $p_0$) reaction at neutron energy 1.8 MeV, were obtained using SRIM \cite{srim} for different gas mixtures. This calculation is based on the Q-value of the $^{26}$Al($n$, $p_0$) reaction, 4.786 MeV, and considering neutron energies up to about 1.8 MeV. Note that the $^{26}$Al($n$, $\alpha_0$) reaction channel will produce alpha particles of energy about 4.8 MeV, considering neutron energies up to about 1.8 MeV and Q-value of 2.97 MeV. The densities for various gas mixtures were obtained from \cite{gas_density}. Fig.~\ref{fig:i3} shows electron drift velocity as a function of the field for various gas mixtures at respective gas pressures and at room temperature conditions, obtained using the Magboltz program \cite{mag}. We plan to use Ne(90\textdiscount) and CF$_4$(10\textdiscount) gas mixtures within the pressure range of 3 - 4 atmospheres as detection gas. The Ne/CF$_4$ gas mixture is chosen as the detection gas as mixtures containing argon and hydrogen had to be omitted to avoid background from neutron scattering and $^{41}$Ar radioactivity. The addition of CF$_4$ gas has many desirable merits viz; high drift velocity,  helpful for fast detectors, and reduced sensitivity to neutron background compared to hydrogenated molecules and in addition their important advantages for use in large volume detectors are non-ﬂammability, and also, they do not form polymers in the avalanches. Additionally, the gas possesses etching properties capable of effectively eliminating existing deposits on electrodes \cite{sauli}. Pure CF$_4$ gas can also be used at low pressure but it might be necessary to apply higher voltages to compensate for lower gain as compared to Ne/CF$_4$ gas \cite{sauli}.

\subsection{Electrostatic calculations }

\begin{figure}[ht!]
\centering
\includegraphics[width=7.110 cm]{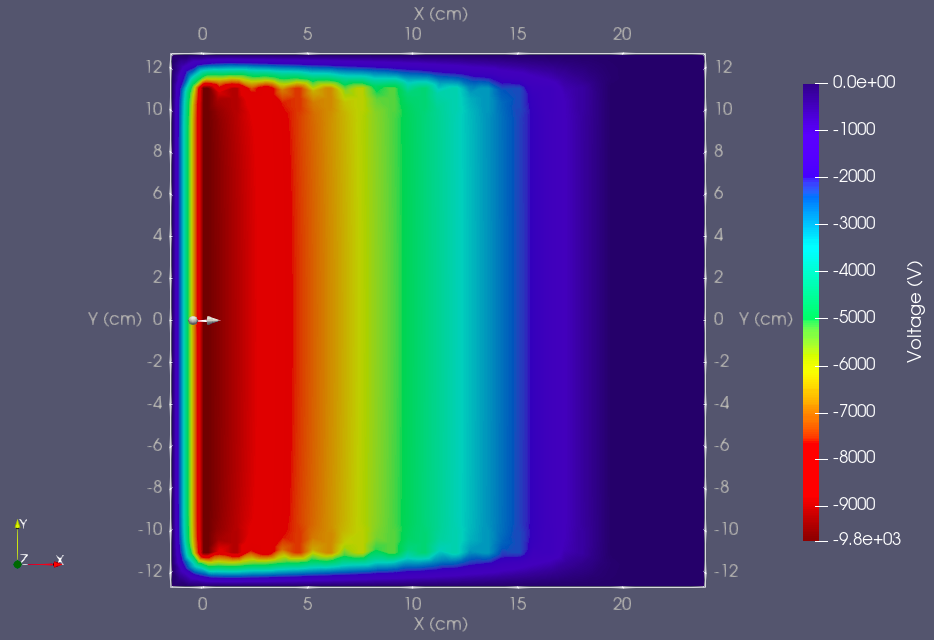}
\qquad
\includegraphics[width=7.0150 cm]{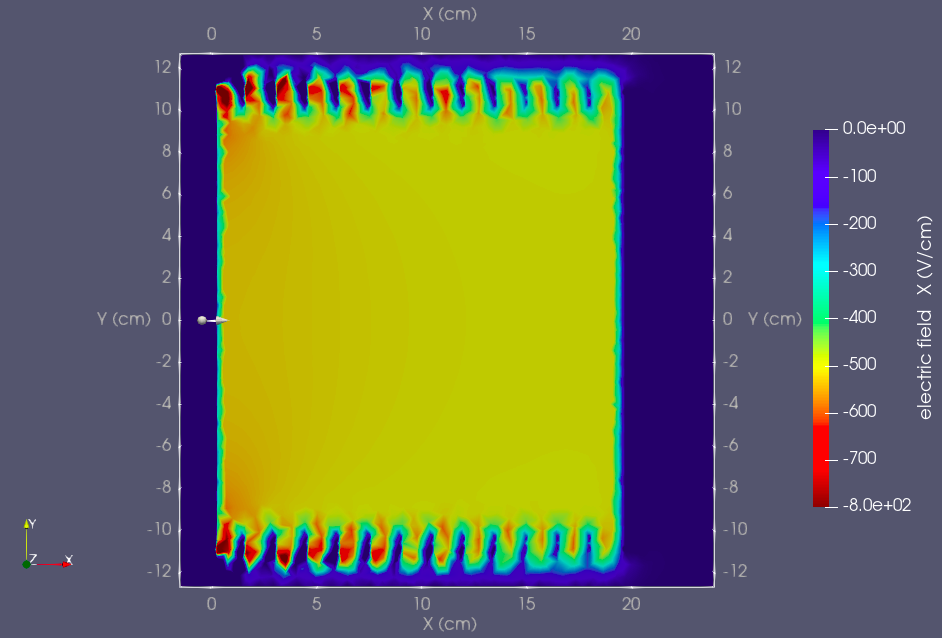}
\caption{(Left) Voltage profile of the detector. (Right) Electric field map of the detector.\label{fig:i5}}
\end{figure}

To estimate the level of electric field uniformity between the cathode, field rings, and anode, an electrostatic field simulation using finite element analysis code Elmer \cite{elmer} is performed. The CAD drawing of the proposed geometry of the detection set-up is opened in Gmsh finite-element mesh-generating code \cite{Gmsh, Gmsh1} and was 3D meshed. The exported mesh file from Gmsh may then be imported into the Elmer graphical user interface, ElmerGUI \cite{elmer}, where the electrostatic problem is defined by specifying materials properties in each sub-volume and voltages can then be specified for all boundary conditions appropriate to the simulation. The full 3D electrostatic calculations were performed for the proposed geometry of the active element of the detector that is the cathode, field rings, and anode.  The final solutions generated from Elmer were read by ParaView \cite{paraview}. Illustrations of the potential profile and electric field map over a cross-section through the $Z$-plane of the geometry as obtained from  ParaView are presented in  Fig.~\ref{fig:i5}.  The left panel of Fig.~\ref{fig:i5} shows the potential (voltage) map of the detector and the right panel of the figure shows the electric field map. The computed field appears to be steady inside the active volume of the detector indicating a stable drift field for particle detection and tracking.

\subsection{Geant4 simulation and background }

\begin{figure}[ht!]
\centering
\includegraphics[width=8.0 cm]{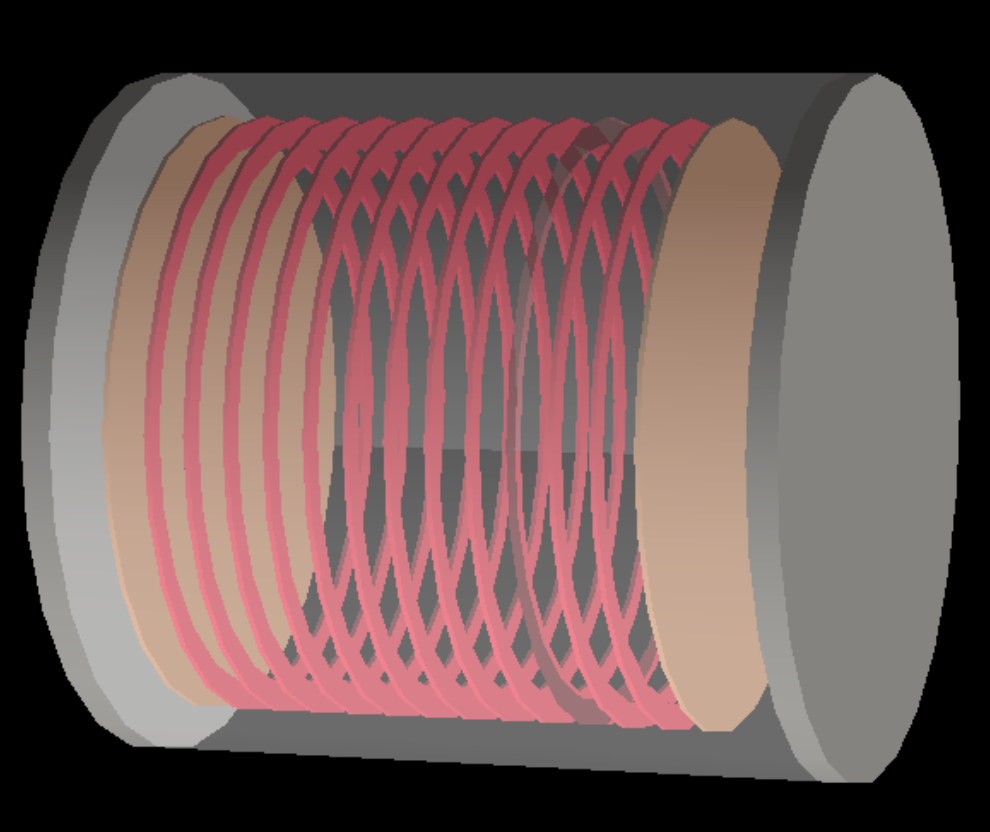}
\includegraphics[width=9.40 cm]{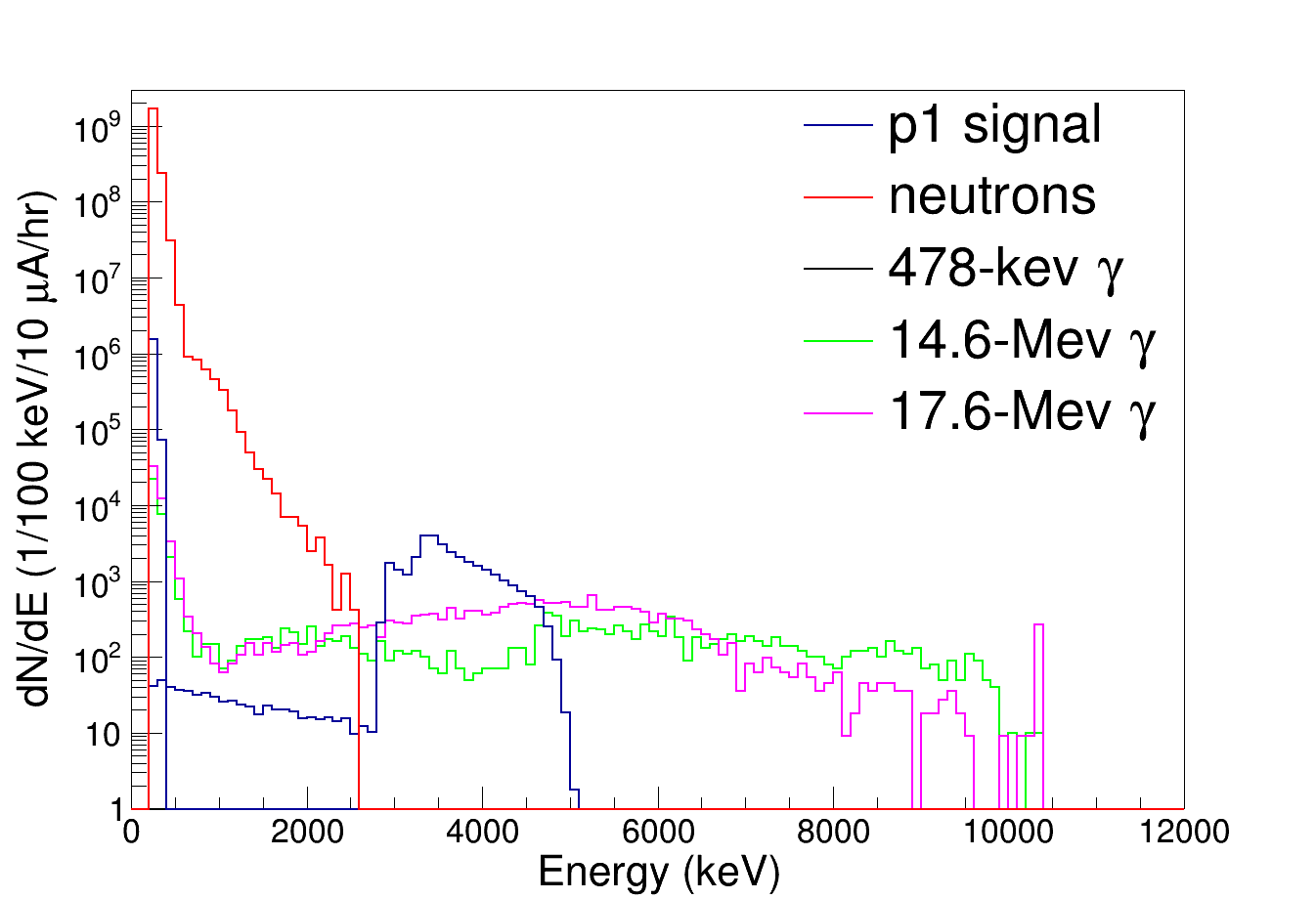}
\caption{(Top) Visualization of the detector chamber as implemented in Geant4 simulation. (Bottom) Simulated background for proton energy of 3.6 MeV, compared to the expected yields from the $^{26}$Al($n$, $p_1$) channel. \label{fig:i6}}
\end{figure}

Geant4 \cite{g4} (for GEometry ANd Tracking ) is a powerful, open-source toolkit for the design and optimization of experimental set-ups used in various scientific fields.  In order to investigate the feasibility of the experimental set-up and expected background level and improve the design of the detector, we carried out a Geant4  simulation of the designed experimental set-up as presented in Fig.~\ref{fig:i}.  The visualization of the experimental set-up as implemented in the simulation is presented in the top panel of Fig.~\ref{fig:i6}.  One of the significant challenges of this experiment is the intense expected background. The primary background sources are neutron and $\gamma$ radiation. The neutron spectrum will mainly produce low-energy background due to hadron scattering in the gas. Similarly, an intense 478-keV $\gamma$ radiation from the $^{7}$Li($p$, $p$\textasciiacute $\gamma$) reaction will cause low-energy background due to Compton scattering in the gas. This background can be reduced by adding a 1.6 cm lead shielding between the source and the detector. Other $\gamma$ radiations background are the 14.6 and 17.6 MeV $\gamma$-rays from the $^{7}$Li($p$, $\gamma$) reaction. Although they are of lower intensity and have a low probability of interacting with the gas, they still contribute via ($\gamma$, p) and ($\gamma$, $\alpha$) reactions. The  $\beta^+$ decays from the target itself do not contribute much to the background.   

The bottom panel of Fig.~\ref{fig:i6} shows the estimated background from all sources as calculated by Geant4 simulation for 3.6 MeV incident proton energy on Li along with expected yields and energies from the $^{26}$Al($n$, $p_1$) channel, which is the most important channel for the total $^{26}$Al$(n, p)$ rate. The intensity of the 478-keV gammas was calculated using the $^{7}$Li($p$, $p$\textasciiacute $\gamma$) cross sections from Presser and Bass \cite{pb}, and Mateus \textit{et al.} \cite{mateus}. The intensities of the 14.6 and 17.6 MeV $\gamma$-rays are based on Tessler \textit{et al.} \cite{tessler} and are scaled to amount to the higher beam energy. The simulation’s statistics were equivalent to a 60-minute run at a beam intensity of 10 $\mu$A and proton energy of 3.6 MeV, which result in the worst background level.  The broad proton peak is due to the wide neutron energy distribution. The low energy tail of the proton signal originates from the backscattering of protons from the target backing. Another issue is the background rate. Although it is possible to define a high threshold for this experiment and reduce the data acquisition load, the actual ionization rate in the gas might hinder the operation of the detector. One of the goals of the segmentation on the anode plane we chose is to reduce the rate on a specific pad to a reasonable level. Based on energy deposition on the trigger pad, a threshold value can be set. Considering the chamber's length of 19.5 cm, we can assume the mean path of a background particle to be 10 cm. A drift velocity of 5 cm/$\mu$s, and an active region of length 19.5 cm, will result in  $\approx$2$\mu$s dead time per event. We calculated from the simulation the event rate for each pad and for each background particle separately, and we obtained, for the highest proton beam, for the central pad, a trigger rate of $\approx$ 200 kHz, leading to a dead time of $\approx$ 40\%. For the outer pads lying on the next 2 radii, which also serve as triggering pads, we obtained a much lower rate of $\approx$ 75 kHz, resulting in a dead time of $\approx$ 15\%. While the dead time on the central pad is relatively high, we do not anticipate it to significantly impact our experiment. First, we can address this issue by segmenting the central pad similarly to the other outer rings, which will reduce the triggering rate by a factor of 4, leading to a manageable dead time. Second, since our primary interest is counting the number of $(n,p)$ events, even if we neglect the energy deposition of the proton on the central pad during its dead time, we might observe a broadening in the energy spectrum, but we will still be able to identify the proton and count the event. Finally, the high triggering rate and large dead time observed in the simulation were specific to the highest energy proton beam used, resulting in elevated background rates. When conducting measurements at lower proton beam energies, the intensity of neutrons and $\gamma$-radiation, which are the primary contributors to the triggering rates, will significantly decrease. It is important to emphasize that the rates discussed here are distinct from the readout rates. The readout rates can be significantly reduced by orders of magnitude by implementing a total energy trigger.

To estimate the detector potential for the case of $^{26}$Al, we consider a 100-nCi target and use the TENDL-2019 cross-section evaluation to calculate the proton and alpha production rates. The estimated statistics and signal-to-background ratio indicate that the statistical uncertainties will be pushed well below the systematic uncertainties which are expected to be on the order of 20\%-30\% due to the differences between the reconstructed spectra and a Maxwellian-flux distribution (See Ref. \cite{MF}). Currently, there are no experimental rates for the $^{26}$Al$(n,\alpha)$ reaction above 1 GK, and the uncertainties for the  $^{26}$Al$(n,p)$ are on the order of factor 3 or even higher (See Ref. \cite{Illiadis}). Hence, a 30\% uncertainty will suffice to substantially constrain those reaction rates.

\section{Summary and Conclusion }
\label{sec:summary}

Measurements of neutron-induced charge particle reaction cross-sections on unstable nuclei at explosive stellar temperatures require high-intensity neutron fluxes at the relevant energies. We plan to use the $^{7}$Li$(p, n)$$^{7}$Be reaction with varying proton energies to produce a set of cross-section measurements that can be combined with proper weights to obtain Maxwellian-Averaged Cross-Section (MACS) values at temperatures of 1.5–3.5 GK. The design and simulation study of the experimental set-up for the neutron-induced charge particle reaction studies for nuclei of astrophysics importance is presented in this work. The results of the simulation show that the experimental approach presented in the paper is feasible despite the challenges with regard to the huge background expected and stable operation of the detector can be obtained. The detector is under construction and will be tested,  characterized, and will be commissioned for the first experiment with $^{40}$K$(n, cp)$ reaction cross-section measurement.

\acknowledgments

This work was supported by ISF grant number 2211/22 and the EU Horizon 2020 Grant No. 101008324 ChETEC-INFRA.

\end{document}